\documentclass[fleqn,10pt]{wlscirep}
\usepackage[utf8]{inputenc}
\usepackage[T1]{fontenc}
\title{Tilting light's polarization plane to spatially separate the nonlinear optical response of chiral molecules on ultrafast timescales}

\author[1,2,*]{Laura Rego}
\author[2,3,+]{David Ayuso}

\affil[1]{Department of Physics, Imperial College London, SW7 2AZ London, United Kingdom}
\affil[2]{Universidad de Salamanca, 37008 Salamanca, Spain}
\affil[3]{Max-Born-Institut, 12489 Berlin, Germany}

\affil[*]{laura.rego@imperial.ac.uk}
\affil[+]{david.ayuso@imperial.ac.uk}

\begin{abstract}
Distinguishing between the left- and right-handed versions of a chiral molecule (enantiomers) is  vital, but also inherently difficult.
Traditional optical methods using elliptically or circularly polarized light rely on weak linear effects which arise beyond the electric-dipole approximation, posing major limitations for time resolving ultrafast chiral molecular dynamics.
Here we show how, by tilting the plane of polarization of an ultrashort burst of intense elliptically polarized light, towards its propagation direction, 
we can turn the light field into a highly efficient chiro-optical tool.   
This ``forward tilting'' can be achieved by focusing the beam tightly, creating structured light which exhibits a nontrivial polarization pattern in space. 
We demonstrate that our structured field allows us to realize an interferometer for efficient chiral recognition that separates the nonlinear optical response of left- and right-handed molecules in space. 
Our work provides a simple, yet highly efficient, way of spatially structuring the polarization of light to image molecular chirality, with extreme enantio-sensitivity and on ultrafast time scales.
\end{abstract}

\begin{document}

\flushbottom
\maketitle

\thispagestyle{empty}

Chirality plays key roles in nature, from dictating the behaviour of subatomic particles \cite{Starosta2001PRL}
to selecting the mating partners of snails \cite{Richards2017EvLett}.
In general, an object is chiral when it cannot be superimposed to its mirror image, with our hands being the typical example.
In chemistry, the left- and right-handed versions of a chiral molecule are called enantiomers.
Their handedness is essential in molecular recognition, and thus distinguishing between opposite enantiomers is vital in many different fields, including bio-medicine, organic chemistry or materials science.
However, chiral distinction is challenging, as opposite molecular enantiomers behave identically unless they interact with another chiral object, such as another chiral molecule or chiral light.

Cutting-edge laser technology creates exciting opportunities for studying molecular chirality, allowing us to access the natural temporal and spatial scales of molecules with unprecedented sub-femtosecond and sub-Angstrom resolution \cite{Krausz2009RMP}.
Important breakthroughs include the real-time observation of electronic currents in atoms \cite{Remetter2006NatPhys,Uiberacker2007Nat,Eckart2018NatPhys},
 molecules \cite{Calegari2014Science,Kraus2015Science,Mansson2021CommChem},
 and solids \cite{Kruger2011Nat,Silva2019NatPhoton,Hui2022NatPhoton}.
 Yet, despite these groundbreaking achievements, imaging the three-dimensional chiral currents governing enantio-sensitive chemical reactions is still very challenging, as natural chiral light is ill-suited for this purpose.

Circularly polarized light is a standard tool for chiral recognition.
In photo-absorption circular dichroism, one measures the differential absorption of left- and right-handed circularly polarized photons, $\Delta I = I_L - I_R$, in a chiral medium \cite{Barron2004}.
While $\Delta I$ has opposite sign in opposite in opposite molecular enantiomers, it is only a small fraction of the total intensity of the optical response ($I_L \simeq I_R$), making chiral recognition challenging, especially on ultrafast time scales.
The reason behind this weak enantio-sensitivity is that circularly polarized light owes its handedness to the (chiral) helix that the tip of the electric-field vector draws in space.
The pitch of this helix, determined by the light's wavelength, is usually orders of magnitude larger than the molecules, particularly in small- to medium-size molecules, which are of especial interest in biochemistry, unless one uses short-wavelength radiation such as X-rays \cite{Zhang2017ChemSci}, or high laser intensities to drive chiral high harmonic generation (HHG) \cite{Cireasa2015NatPhys,Smirnova2015JPB,Ayuso2018JPB,Ayuso2018JPB_model,Harada2018PRA,Baykusheva2018PRX}.

This limitation can be bypassed by creating synthetic chiral light \cite{Ayuso2019NatPhot,Ayuso2021NatComm,Ayuso2022PCCP,Neufeld2021PRR,Katsoulis2021,Mayer2021,Khokhlova2021}, where the tip of the electric-field vector draws a three-dimensional chiral Lissajous figure in time.
The enantio-sensitive response of chiral media to such locally chiral light is driven by purely electric-dipole interactions, and it is orders of magnitude stronger than in traditional optical methods.
Another strategy is stop relying on light's chirality \cite{Ordonez2018PRA,Ayuso2022persp} and record enantio-sensitive signals by analysing vectorial \cite{Ritchie1976PRA,Powis2000JCP,Bowering2001PRL,Garcia2003JCP,Lux2012Angew,Stefan2013JCP,Garcia2013NatComm,Janssen2014PCCP,Lux2015ChemPhysChem,Kastner2016CPC,Comby2016JPCL,Beaulieu2016FD,Beaulieu2017Science,Beaulieu2018NatPhys,Comby2018NatComm,Fischer2000PRL,Belkin2001PRL,Fischer2002CPL,Patterson2013Nat,Neufeld2019PRX,Ayuso2022OptExp,Owens2018PRL,Yachmenev2019PRL,Milner2019PRL} or tensorial \cite{Demekhin2018PRL,Goetz2019PRL,Rozen2019PRX,Ordonez2022PCCP} observables, such as the direction of the photoelectron current upon ionization with circularly \cite{Ritchie1976PRA,Powis2000JCP,Bowering2001PRL,Garcia2003JCP,Lux2012Angew,Stefan2013JCP,Garcia2013NatComm,Janssen2014PCCP,Lux2015ChemPhysChem,Kastner2016CPC,Comby2016JPCL,Beaulieu2016FD,Beaulieu2017Science,Beaulieu2018NatPhys} or elliptically \cite{Comby2018NatComm} polarized light.
This asymmetric current is orthogonal to the plane of polarization of the wave, it has opposite directions in opposite molecular enantiomers, and it is strong because it is driven by purely electric-dipole interactions.
However, if one seeks to induce and record an equivalent current via nonlinear excitation with elliptically polarized light using an all-optical setup, measuring the radiation emitted by the oscillations of the induced polarization in time, they will encounter a fundamental limitation: while the generation of an asymmetric current is symmetry-allowed, it is not in the right direction.

Let us analyze the highly nonlinear response of a medium of randomly oriented chiral molecules to an elliptically polarized driving field $\textbf{E}$ propagating in the $\hat{\bold{z}}$ direction within the electric-dipole approximation,
\begin{equation}\label{eq:E}
\textbf{E}(t) = E_0 \, a(t) [\cos(\omega t + \phi_{\text{CEP}}) \hat{\bold{x}}+ \varepsilon \sin(\omega t + \phi_{\text{CEP}}) \hat{\bold{e}}_\varepsilon],
\end{equation}
where $E_0$ is the field amplitude,
$a(t)$ is the temporal envelope,
$\phi_{\text{CEP}}$ is the carrier-envelope phase (CEP),
and $\varepsilon<1$ is the ellipticity.
In a standard elliptical wave, the unitary vector defining the direction of the minor ellipticity component is simply $\hat{\bold{e}}_\varepsilon=\hat{\bold{y}}$, i.e. the polarization plane is orthogonal to the propagation direction $\hat{\bold{z}}$.
The nonlinear polarization induced by this field in isotropic chiral media can have three symmetry-allowed orthogonal polarization components \cite{Ayuso2022OptExp}:
\begin{equation}\label{eq:P}
\textbf{P}(t) = P_0(t) \hat{\bold{x}} + P_\varepsilon(t) \hat{\bold{e}}_\varepsilon + P_c^{L/R}(t) \hat{\bold{e}}_c,
\end{equation}
where $\hat{\bold{e}}_c = \hat{\bold{x}} \times \hat{\bold{e}}_\varepsilon$.
The in-plane components $P_0$ and $P_\varepsilon$ are \emph{achiral}:
they are identical in left- and right-handed molecules.
The out-of-plane component $P_c^{L/R}$ is \emph{chiral}: it is exclusive of chiral media, and it has equal intensity and opposite phase in opposite molecular enantiomers, $P_c^L=-P_c^R$.
However, this  component is completely invisible in the typical macroscopic HHG signal, which can only record the (achiral) components which are orthogonal to the propagation direction $\hat{\bold{z}}$.
For this reason, the recent chiral HHG experiments with elliptical \cite{Cireasa2015NatPhys} and two-colour \cite{Harada2018PRA,Baykusheva2018PRX} driving fields relied on effects which arise beyond the electric-dipole approximation, limiting the enantio-sensitivity of the HHG camera.

\begin{figure}[h]
\centering
\includegraphics[width=\textwidth]{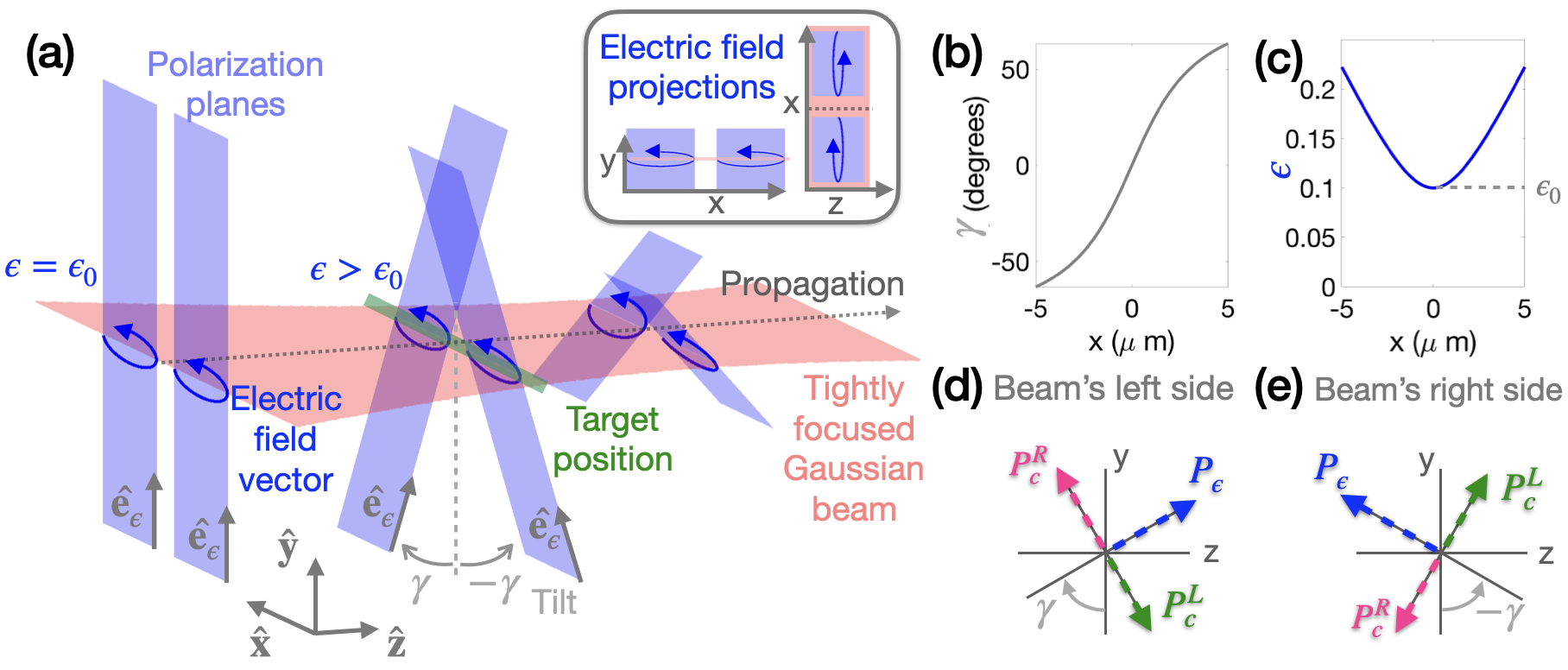}
\caption{\label{fig:1} \textbf{Tilting the plane of light's polarization.} \textbf{a,} A laser field with elliptical polarization (blue arrows and planes) and a Gaussian profile (pink) acquires a forward polarization tilt upon tight focusing.
The tilt angle $\gamma$ of the minor ellipticity component $\hat{\bold{e}}_\varepsilon$ is opposite at opposite sides of the beam's axis.
The inset shows the projections of the electric field polarization on the $yx$ and $xz$ planes.
\textbf{b,c,} Tilt angle $\gamma$ (\textbf{b}) and total ellipticity (\textbf{c}) as functions of the transverse coordinate $x$ for a beam's waist of $W=2.5 \mu m$.
\textbf{d,e,} Schematic representation of the polarization induced in randomly oriented chiral molecules at each side of the beam's axis, see Eq. \ref{eq:P}.
Note that the chiral component of the induced polarization has opposite orientation in opposite molecular enantiomers.}
\end{figure}

 Here we show that this fundamental limitation can be overcome by tilting the plane of polarization of the wave towards is propagation direction.
Such a ``forward tilt" can be achieved by focusing the laser beam tightly, creating light with structured polarization in space, see Fig. \ref{fig:1}.
Our proposal exploits the potential of structuring light's polarization \cite{Maurer2007NJP,Zhan2009Optica,Beckley2010Optica,Pisanty2019NatPhoton,Pisanty2019PRL,Dorney2019NatPhoton,Rego2022SciAdv} to realize an enantio-sensitive interferometer for efficient chiral recognition that separates the nonlinear response of opposite molecular enantiomers in space.

We consider an ultrashort and tightly focused Gaussian beam with elliptical polarization, where the pulse duration is only a few cycles, and the size of the beam waist is only a few times larger than its wavelength. 
Such tight focusing creates a strong longitudinal electric-field component \cite{Bliokh2015}, along the propagation direction, see Fig. \ref{fig:1}a.
The consequences of this new component are twofold.
First, the plane of polarization of the electric-field vector rotates around the $x$ axis, in opposite directions at opposite sides of the beam propagation axis.
Second, the ellipticity increases.
Note that both the tilt angle $\gamma$ and the ellipticity $\varepsilon$ become spatially structured, with $\gamma\,(-x)=-\gamma\,(x)$ and $\varepsilon\,(-x)=\varepsilon\,(x)$, see Figs. \ref{fig:1}b,c and Supplementary Information.
This ``forward tilt" is a key aspect of our proposal:
the direction of the minor ellipticity component of the driving field $\hat{\bold{e}}_\varepsilon$ (see Eq. \ref{eq:E}), is no longer orthogonal to the propagation direction $\hat{\bold{z}}$ ($\hat{\bold{e}}_\varepsilon\neq\hat{\bold{y}}$), see Fig. \ref{fig:1}a.
As we show in the following, the rotation of the polarization plane with respect to the propagation direction allows us to realize an efficient interferometer for chiral recognition.

The induced polarization driven by our tilted laser field in opposite molecular enantiomers is depicted in Fig. \ref{fig:1}d,e.
In the laser reference frame, defined by the two laser polarization vectors ($\hat{\bold{x}}$ and $\hat{\bold{e}}_\varepsilon$) and $\hat{\bold{e}}_c=\hat{\bold{x}}\times\hat{\bold{e}}_\varepsilon$, the three components of the induced polarization have exactly the same intensity in opposite enantiomers, and thus $\lvert\textbf{P}\rvert_L=\lvert\textbf{P}\rvert_R$.
The only enantio-sensitive quantity is the direction of the chiral polarization component $P_c^L=-P_c^R$.
To create an enantio-sensitive intensity, one would need to project $P_c^{L/R}$ and one of the two of achiral components ($P_0$ and $P_\varepsilon$) over a common polarization axis that is in the right direction to produce a phase-matched macroscopic signal.
This is exactly what the proposed optical setup does:
by tilting the plane of polarization of light, we project $P_c$ and $P_\varepsilon$ over the common $y$ axis, where they interfere:
\begin{equation}\label{eq:Py}
P_y(t) = [ P_\varepsilon(t) \hat{\bold{e}}_\varepsilon + P_c^{L/R}(t) \hat{\bold{e}}_c ] \cdot \hat{\bold{y}}
= P_\varepsilon(t) \cos(\gamma) + P_c^{L/R}(t) \sin(\gamma).
\end{equation}
Eq. \ref{eq:Py} shows that the tilt angle $\gamma$ and the molecular handedness ($P_c^L=-P_c^R$) control the relative sign between the achiral and chiral components of the induced polarization, creating an enantio-sensitive interferometer. 
Since $\gamma\,(x)=-\gamma\,(-x)$, the chiral ``arm'' of our interferometer, $P_c^{L/R} \sin(\gamma)$, has opposite phase at opposite sides of the beam axis.
As a result, the effect of the spatial displacement $x\leftrightarrow -x$ is equivalent to reversing the molecular handedness ($L\leftrightarrow R$). 

We note that, in a long laser pulse, with at least several optical cycles of similar amplitude driving HHG, the chiral polarization component $P_c^{L/R}$ can only contain even harmonic frequencies \cite{Ayuso2019NatPhot}, whereas the achiral components $P_0$ and $P_\varepsilon$ carry odd harmonic frequencies.
However, these selection rules relax as we reduce the pulse duration, and the chiral and achiral components can spectrally overlap \cite{Ayuso2021Optica}.
Here we take advantage of the broad spectral bandwidth of ultrashort laser pulses to realize an enantio-sensitive interferometer.

\begin{figure}[h!]
 \centering
\includegraphics[width=1\textwidth]{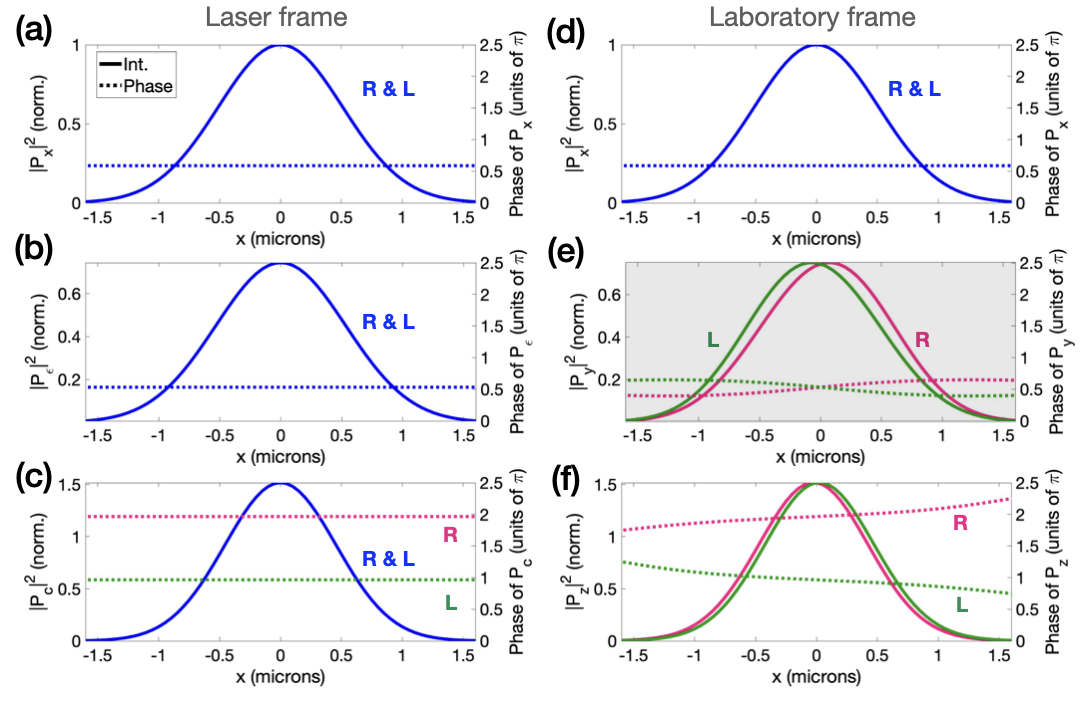}
\caption{\label{fig:2} \textbf{Nonlinear response of randomly oriented propylene oxide in the near field.}
Intensity (solid lines) and phase (dotted lines) at the $6^{th}$-order harmonic frequency in laser (\textbf{a-c}) and laboratory (\textbf{d-f}) reference frames (see main text) for the right- (pink) and left-handed (green) enantiomers (the non-enantio-sensitive curves are in blue).
The intensity is symmetric with respect to $x$ and not enantio-sensitive in the laser reference frame (\textbf{a-c}), but it becomes asymmetric and enantio-sensitive when projecting $P_\varepsilon$ and $P_c$ over the laboratory-frame axes $y$ and $z$.
The enantio-sensitive component which can be detected in the macroscopic far-field signal $P_y$ is highlighted in grey shading.
Laser parameters: $\varepsilon_0=0.1$, $I=6 \times 10^{13} \, W cm^{-2}$, focal diameter 5$\mu$m, wavelength $\lambda$=780 nm, pulse duration 7 fs (FWHM) and $\phi_{CEP}=\pi/4$.}
\end{figure}

We have modelled the ultrafast electronic response of randomly oriented propylene oxide molecules to the structured laser field presented in Fig. \ref{fig:1} using a state-of-the-art implementation of real-time time-dependent density functional theory \cite{Tancogne2020,Andrade2015,Castro2006,Marques2003}, see Methods.
We have considered the following laser parameters: peak intensity $I=6 \cdot 10^{13} \,$ W$/$cm$^{2}$, incoming ellipticity $\varepsilon_0=0.1$, focal diameter 5$\mu$m, central wavelength of $\lambda$=780 nm and 7 fs of pulse duration (FWHM).
Fig. \ref{fig:2} shows the amplitude and phase profiles of the induced polarization at frequency $6\omega$ when using $\phi_{CEP}=\pi/4$, in the laser $\{\hat{\bold{x}},\hat{\bold{e}}_\varepsilon,\hat{\bold{e}}_c=\hat{\bold{x}}\times\hat{\bold{e}}_\varepsilon\}$ (Fig. \ref{fig:2}a-c) and laboratory $\{\hat{\bold{x}}, \hat{\bold{y}}, \hat{\bold{z}}\}$ (Fig. \ref{fig:2}d-e) reference frames.

In the laser reference frame (Fig. \ref{fig:2}a-c), the three components of the induced polarization have the same intensity in opposite enantiomers.
The molecular handedness is encode in the phase $P_c^{L/R}$, see Fig. \ref{fig:2}c.
Thanks to the tilt of the polarization plane, when projecting $P_c^{L/R}$ and $P_\varepsilon$ on the laboratory-frame vectors, $\lvert P_y \rvert^2$ and $\lvert P_z \rvert^2$ become asymmetric with respect to the propagation axis and enantio-sensitive,
see Fig. \ref{fig:2}d-f.
Note that the two reference frames are rotated around the $x$ axis, and thus induced polarization in this direction ($P_x$) is identical in both frames.

We now take advantage of what used to be a fundamental limitation: the fact that only $P_x$ and $P_y$ can generate a phase-matched HHG signal ($P_z$ is invisible in the macroscopic response).
By tilting the plane of polarization of light, we make $P_y$ and $P_z$ enantio-sensitive, and thus the intensity of phase-matched harmonic emission, proportional to $\lvert P_x \rvert^2+\lvert P_y \rvert^2$, becomes enantio-sensitive.
That is, we use the propagation vector of light $\bold{k}\propto\hat{\bold{z}}$ as a ``filter'', to control which components of the induced polarization are observed in the macroscopic HHG signal and which cannot.

\begin{figure}[h!]
\centering
\includegraphics[width=1\textwidth]{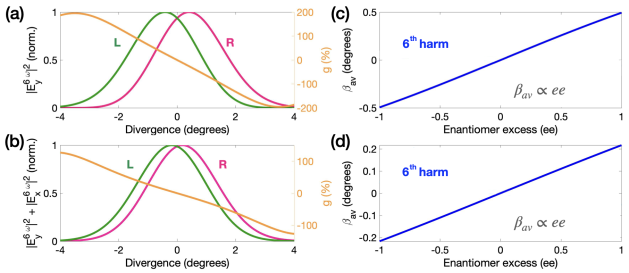}
\caption{\label{fig:3} \textbf{Enantio-sensitive HHG in the far field.}
\textbf{a,b} Intensity associated with the $y$-polarized component (\textbf{a}) of the radiation emitted from right- (pink) and left-handed (green) propylene and total intensity (\textbf{b}) at frequency $6\omega$, and dissymmetry factor (orange).
\textbf{c,d} Average divergence angle in the intensity of the $y$-polarized component (\textbf{c}) and in total intensity (\textbf{d}) as functions of the enantiomeric excess.
See caption of Fig. \ref{fig:2} for laser parameters.}
\end{figure}

Fig. \ref{fig:2}e shows that the intensity profile of the $y$-polarized component of the induced polarization is asymmetric and different for left- and right-handed molecules.
However, the relevant asymmetry to realize an enantio-sensitive interferometer in the far field is not the near-field intensity, but its phase, which is also enantio-sensitive.
As shown in Fig. \ref{fig:2}e, the phase of the nonlinear response at frequency $6\omega$ increases with $x$ in the right-handed molecules, and it decreases in the left-handed molecules.
This behaviour is similar for other harmonic frequencies (not shown).
The asymmetric and enantio-sensitive phase profile shown in Fig. \ref{fig:2}e determines the propagation direction of the emitted harmonic light.
For this choice of parameters, the left-handed molecules emit harmonic light preferentially to the left, whereas the right-handed molecules radiate preferentially to the right, see Fig. \ref{fig:3}.

The direction of harmonic emission is strongly enantio-sensitive when we consider the intensity associated with the $y$-polarized component of the emitted harmonic light $\lvert E_y \rvert^2$, which is generated by $P_y$.
This polarization component could be separated from the non-enantio-sensitive component $\lvert E_x \rvert^2$ by placing a polarizer before the detector.
However, our enantio-sensitive observable remains strong when considering the total intensity of harmonic emission, proportional to $\lvert E_x \rvert^2 + \lvert E_y \rvert^2$, as shown in Fig. \ref{fig:3}b.

To quantify the degree of enantio-sensitivity in the macroscopic far-field signal, we use an angularly resolved definition of the dissymmetry factor $g=2 \,(I_L-I_R)/(I_L+I_R)$, where $I_{L/R}$ is the intensity of harmonic light emitted from the left-/right-handed enantiomer at a given divergence angle.
The values of the dissymmetry factor approach the ultimate efficiency limit ($\pm 200\%$) in the $y$-polarized component of the emitted harmonic light (Fig. \ref{fig:3}c), but are also very strong when we consider the total macroscopic intensity (Fig. \ref{fig:3}d).

The proposed optical setup allows us to unequivocally determine the relative concentration opposite enantiomers in mixtures, which is usually quantified via the enantiomeric excess, $ee=(C_R-C_L)/(C_R+C_L)$, where $C_{L/R}$ is the concentration of left-/right-handed molecules.
The average divergence angle in the far-field harmonic intensity is approximately proportional to the enantiomeric excess, as shown in Fig. \ref{fig:3}c and Fig. \ref{fig:3}d for the $y$-polarized component and for the total intensity, respectively.
Our numerical simulations show that the average angles of emission for enantio-pure samples are $\pm 0.5^\circ$ when we consider the $y$-polarized component of the harmonic light, and $\pm 0.2^\circ$ in the total intensity, for the parameters and molecule considered in this work.

\begin{figure}[h!]
 \centering
\includegraphics[width=1\textwidth]{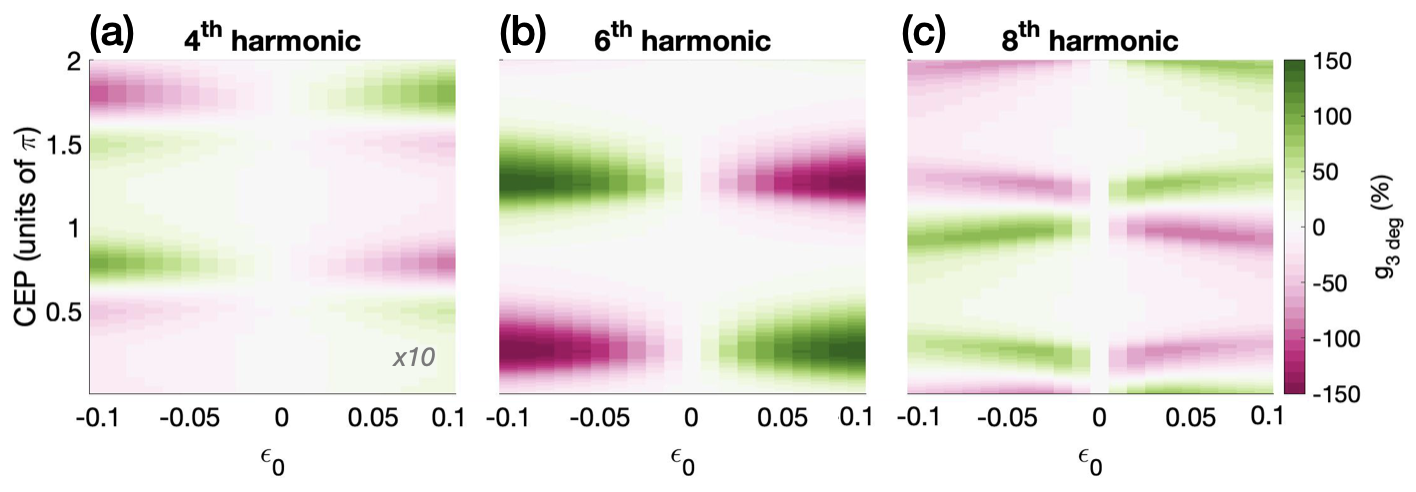}
\caption{\label{fig:4} \textbf{Maximizing the enantio-sensitive response of propylene oxide.}
Dissymmetry factor at a divergence angle of 3 degrees as a function of the CEP and $\varepsilon_0$ at the $4^{th}$ (\textbf{a}), $6^{th}$ (\textbf{b}), and $8^{th}$ (\textbf{c}) harmonic orders.
Reversing the sign of $\varepsilon_0$ results in a change of sign in $g$, and is equivalent to changing the CEP by $\pi$.
See caption of Fig. \ref{fig:2} for laser parameters.}
\end{figure}

We can control the polarization of the structured driving field, and thus the enantio-sensitive response of the chiral molecules, by adjusting the laser parameters in the proposed optical setup.
This allows us to optimize the asymmetry in the direction of HHG emission for each harmonic number.
Fig. \ref{fig:4} shows the dissymmetry factor in the $y$-polarized component of the emitted light at the $4^{th}$ (a), $6^{th}$ (b) and $8^{th}$ (c) harmonic frequencies at a divergence angle of 3 degrees, as functions of the CEP and $\varepsilon_0$.
The values of the CEP and $\varepsilon_0$ which maximize the enantio-sensitive response of the molecules are different for different harmonic numbers, reflecting the fact that the relative amplitude and phase between the achiral ($P_\varepsilon$) and chiral ($P_c^{L/R}$) components of the induced polarization are frequency-dependent.

The phase of the chiral component of our interferometer $P_c^{L/R}\sin(\gamma)$ (see Eq. \ref{eq:Py}) depends on the relative phase between the strong-field component of the laser field and its longitudinal component, which is opposite at opposite sides of the beam axis, and locked to the direction of light propagation \cite{Bliokh2015}.
This means that, while the amplitude of $P_c^{L/R}\sin(\gamma)$ can be controlled by controlling the beam waist of the Gaussian laser beam (controlling the amplitude of the longitudinal field component), we do not have full control over its phase.
However, we can control both the amplitude and phase of the achiral component of the interferometer $P_\varepsilon \cos(\gamma)$ by controlling the CEP of the laser field and the incoming ellipticity, achieving full control over the enantio-sensitive interference, as shown in Fig. \ref{fig:4}.
Note that changing the CEP by $\pi$ changes the sign of $P_\varepsilon$ and thus of the sign of the dissymmetry factor, and that reversing the sign of the incoming ellipticity produces an equivalent effect.

In contrast with previous works \cite{Ayuso2019NatPhot,Ayuso2021NatComm,Ayuso2022PCCP,Neufeld2021PRR,Katsoulis2021,Mayer2021,Khokhlova2021}, the laser field proposed in this work is not locally chiral: the Lissajous figure that the tip of the electric-field vector draws in space is confined to a plane.
The field becomes chiral only when we take into account its propagation direction.
One could think that, in this scenario, the enantio-sensitive response of the chiral molecules must rely on weak magnetic or quadrupole interactions, as it happens in traditional all-optical methods relying on the chirality of elliptically or circularly polarized light \cite{Barron2004}, where the propagation vector plays a key role in defining the wave's handedness, but we have shown that this is not necessarily the case.
Here, the propagation vector plays the role of a ``chiral observer''\cite{Ayuso2022persp}, dictating which components of the induced polarization are allowed to generate a phase-matched radiation that propagates to the detector and which cannot.
It acts as a near-field polarizer, projecting two components of the induced polarization (the chiral component and one of the two achiral components) onto a common axis, where they efficiently interfere thanks the short pulse duration of the driving field.   
As a result of the nontrivial structure of the laser polarization in space, the nonlinear response of the chiral molecules creates an enantio-sensitive wavefront that leads to spatial separation of the radiation emitted from left- and right-handed molecules in the far field.

The proposed optical method realizes an efficient interferometer that allows us to measure the enantiomeric excess in mixtures of left- and right-handed molecules.
Because of the ultrafast nature of the nonlinear interactions responsible of the enantio-sensitive response, our method seems to be ideally suited for monitoring enantio-sensitive chemical reactions in real time, with sub-femtosecond temporal resolution, taking advantage of the well-established time-energy mapping in HHG \cite{Mairesse2003Science,Baker2006Science,Smirnova2009Nature}, which relates the instants of strong-field ionization and radiative recombination to the frequency of the emitted harmonic light.
Furthermore, the enantio-sensitive direction of HHG is a molecule-specific quantity, and thus our proposal creates new exciting opportunities for developing molecular markers of enantio-sensitive chemical dynamics.

\section*{Acknowledgments}
We acknowledge highly stimulating discussions with Olga Smirnova and Misha Ivanov.
D. A. is extremely grateful for their endless and continuous support and guidance.
L. R. acknowledges funding from the European Union-NextGenerationEU and the Spanish Ministry of Universities via her Margarita Salas Fellowship through the University of Salamanca;
D. A. acknowledge funding from the Deutsche Forschungsgemeinschaft SPP 1840 SM 292/5-2;
L. R. and D. A. acknowledge funding from the Royal Society URF$\backslash$R1$\backslash$201333 and RF$\backslash$ERE$\backslash$210358.

\section*{Methods}

\subsection*{Orientational averaging of the laser-induced polarization}
The induced polarization in the medium of randomly oriented propylene oxide molecules was calculated by averaging over the contribution from different molecular orientations:
\begin{equation}
\mathbf{P}(t,x) = \frac{1}{8\pi^2} \int_{0}^{2\pi} \int_{0}^{\pi} \int_{0}^{2\pi} \mathbf{P}_{\phi\theta\chi}(t,x) \; \sin(\theta) \; d\phi \; d\theta \; d\chi,
\end{equation}
where $\phi$, $\theta$ and $\chi$ are the three Euler angles and $\mathbf{P}_{\phi\theta\chi}$ is the polarization response of a particular molecular orientation in the laboratory frame.
We used the Lebedev quadrature \cite{Lebedev1999}
of order 11 (50 points) to integrate over $\phi$ and $\theta$.
For each Lebedev point, the polarization of the electric-field vector of the laser field was defined in a way that its strong-field component pointed in the same direction, which allowed us to reach convergence in $\chi$ using 4 points via trapezoidal numerical integration.

\subsection*{Single-molecule response}
The light-induced polarization was evaluated using real-time time-dependent density functional theory Octopus  \cite{Tancogne2020,Andrade2015,Castro2006,Marques2003}.
We used the local-density approximation \cite{Dirac1930,Bloch1929,Perdew1981} to account for electronic exchange and correlation effects, and the averaged-density self-interaction correction \cite{Legrand2002} to account describe the long-range behaviour of the electron density.
The 1s orbitals of the heavier atoms (carbon and oxygen) are barely affected by the laser field, and they were described using pseudo-potentials.
The Kohn-Sham orbitals and the electron density were expanded into a uniform real-space grid of points separated by $0.4$ a.u. enclosed in a sphere of radius $R=42$ a.u., and we used a complex absorbing potential with width $20$ a.u. and height $-0.2$ a.u. to avoid unphysical reflexions of the electron density.

\subsection*{Far-field image}
The intensity of HHG in the far field was calculated using the Fraunhofer diffraction equation:
\begin{equation}\label{Fraunhofer}
E_\xi(\beta,N) \propto \int_{-\infty}^{\infty} \frac{d^2}{dt^2} P_\xi(x,N) e^{-i\frac{N\omega x}{c\beta}} dx,
\end{equation}
where $\xi=x,y$, $P_x$ and $P_y$ are the non-enantio-sensitive and enantio-sensitive components of the induced polarization as a function of the transverse coordinate $x$ in the near field in the frequency domain,
$N$ is the harmonic number,
$\beta$ is the divergence angle,
$c$ is the speed of light in vacuum, and $\omega$ is the fundamental frequency.

\bibliography{Bibliography}

\newpage

\section*{SUPPLEMENTARY INFORMATION}

\subsection*{Calculation of the forward tilt angle and ellipticity} 

\begin{figure}[h!]
\centering
\includegraphics[width=1\textwidth]{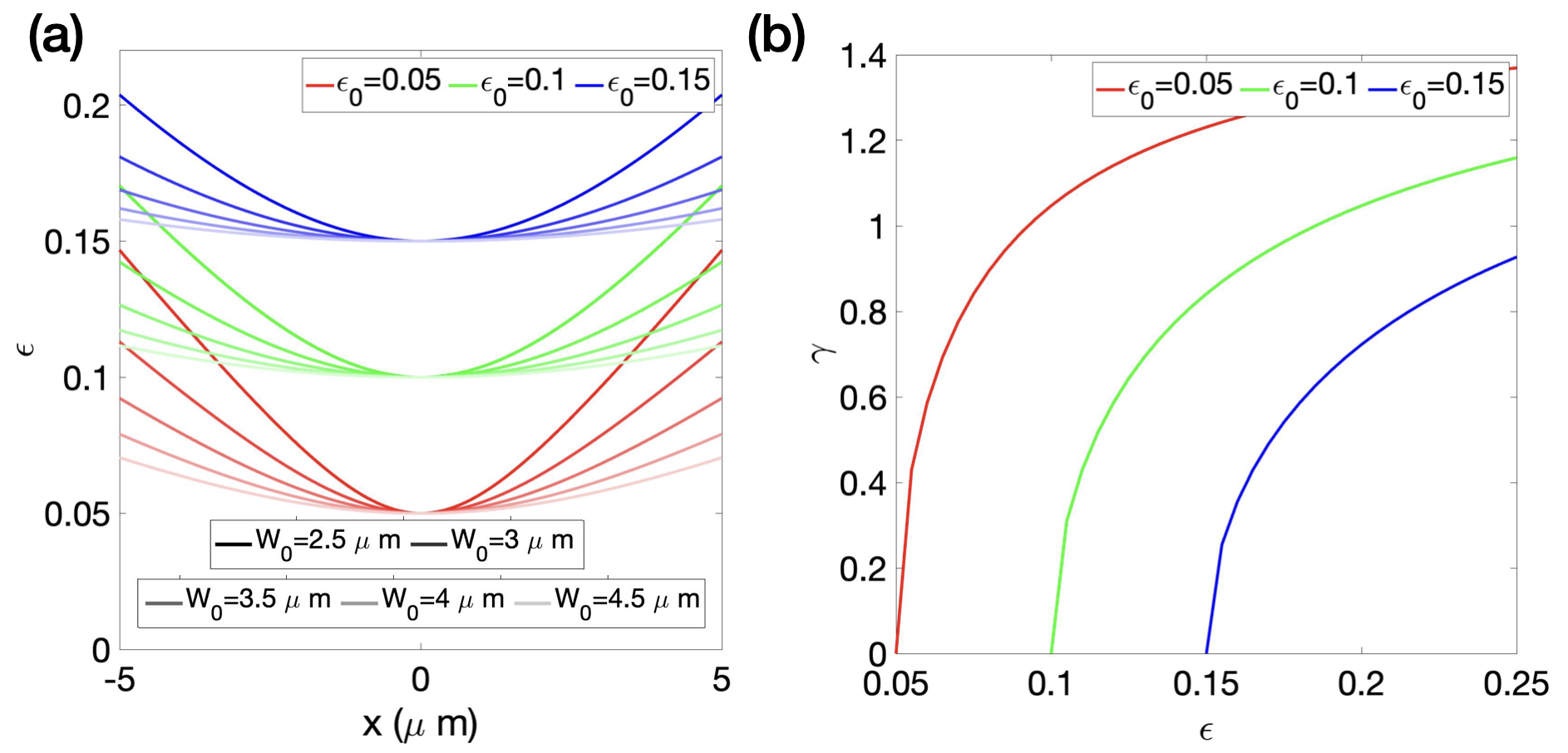}
\caption{\label{fig:S1} 
\textbf{a}, Total ellipticity of the driving beam $\varepsilon$ as a function of the transverse coordinate $x$ for different values of the initial ellipticity $\varepsilon_0$ and the beam waist $W$.
\textbf{b}, Forward tilt angle $\gamma$ as a function of the total ellipticity $\varepsilon$ for different values of the incoming ellipticity $\varepsilon_0$. }
\end{figure}

To calculate the forward tilt angle $\gamma$ and total ellipticity $\varepsilon$ of the driving field upon tight focusing (see Eq. \ref{eq:E}), we write the electric-field vector in terms of the laboratory-frame vectors:
\begin{eqnarray}\label{eq:Exyz}
\textbf{E}(t) = a(t) [E_0 \cos(\omega t + \phi_{\text{CEP}}) \hat{\bold{x}} + (E_y \hat{\bold{y}} + E_z \hat{\bold{z}} ) \sin(\omega t + \phi_{\text{CEP}}) ],
\end{eqnarray}
where $E_{y}=\varepsilon_0 E_{0}$ is the incoming elliptical component, and $E_z$ is the longitudinal component which arises due to tight focusing \cite{Bliokh2015}, $E_z=-2x/(W^2 k) E_{x}$.
The total ellipticity in Eq. \ref{eq:E} and the tilt angle are given by
\begin{align}
\varepsilon &= \frac{\sqrt{E_y^2+E_z^2}}{E_0} = \sqrt{\varepsilon_0^2+[2x/(W^2 k)]^2}, \\
\gamma &= \arctan\bigg(\frac{E_z}{E_y}\bigg) = \arctan{\bigg(\frac{-2 x}{\varepsilon_0 W^2 k}\bigg)}.
\end{align}
To illustrate our degree of control over $\varepsilon$ and $\gamma$, and thus over the structured polarization of our diving field, we present in Fig. \ref{fig:S1} the values of these quantities for different laser parameters.
As shown in Fig. \ref{fig:S1}a, we can control the total ellipticity $\varepsilon$ by adjusting the incoming ellipticity $\varepsilon_0$ and the waist of the gaussian field $W$.
Note that reducing $W$ leads to the generation of a stronger longitudinal component $E_z$, and thus to an increase of $\varepsilon$.
Fig. \ref{fig:S1}b shows the relation between $\gamma$ and $\varepsilon$ for different values of $\varepsilon_0$.
For a given $\varepsilon_0$, higher values of $\varepsilon$ are associated with stronger longitudinal components, and thus with larger $\gamma$.
Note also that increasing $\varepsilon_0$ means increasing $E_y$, and thus reducing $\gamma$.

\subsection*{Calculation of the average dissymmetry factor} 

\begin{figure}[h!]
\centering
\includegraphics[width=1\textwidth]{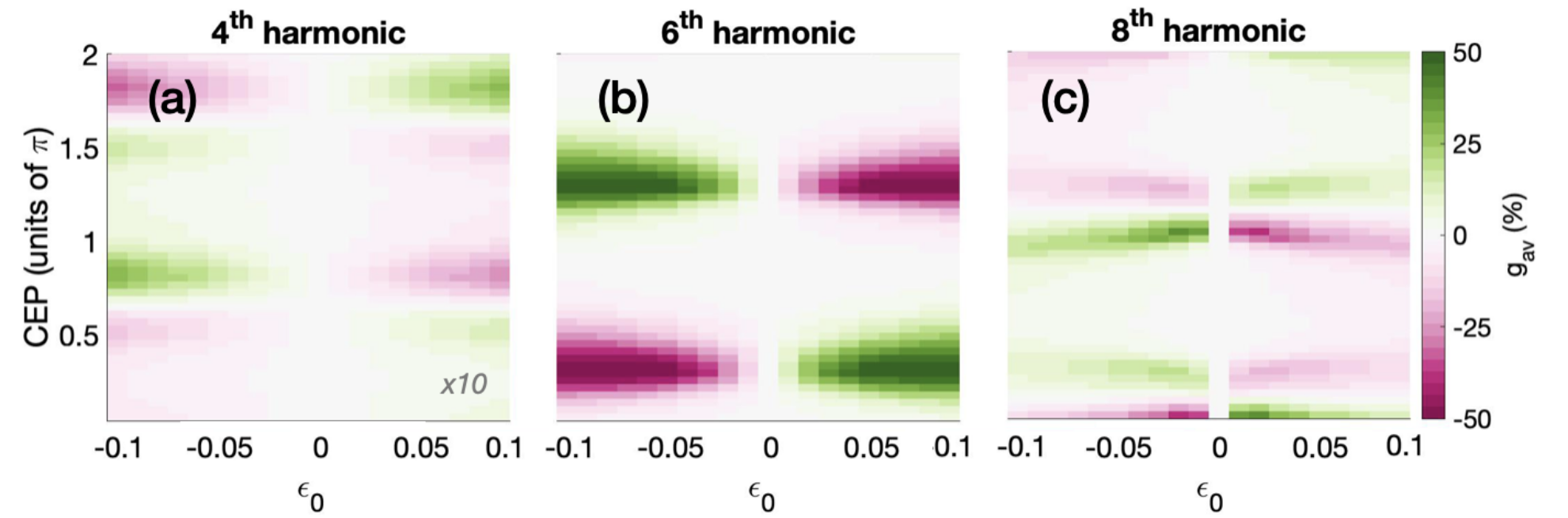}
\caption{\label{fig:S2} Spatially averaged dissymmetry factor as a function of the CEP and the incoming ellipticity $\varepsilon_0$ of the driving laser beam for the $4^{th}$ (\textbf{a}), $6^{th}$ (\textbf{b}), and $8^{th}$ (\textbf{c}) harmonic orders.
The molecule and laser parameters are the same as in Fig. \ref{fig:2} of the main text.}
\end{figure}

The dissymmetry factor $g$ defined in the main is a spatially structured quantity which depends on the divergence angle $\beta$, as shown in Figs. \ref{fig:3}a-b.
We can define an averaged version of this quantity by weighting the $g$ with the total intensity at value of $\beta$.
Since $g$ has opposite signs at each side of the beam's axis (Figs. \ref{fig:3}a-b), we integrate over positive angles only:
\begin{eqnarray}
	 g_{av} & = &\frac{\int_0^{\infty} g (\beta) \, I(\beta) d\beta}{\int_0^{\infty} I(\beta) d\beta}. \label{eq:gav}
\end{eqnarray}
Fig. \ref{fig:S2} shows the averaged dissymmetry factor for the (a) $4^{th}$, (b) $6^{th}$, and (c) $8^{th}$ harmonic orders, as a function of the CEP and the incoming ellipticity $\varepsilon_0$.
Our results show that the enantio-sensitive response remains strong upon integration over the emission angle, and that the values of $\phi_{CEP}$ and $\varepsilon$ that maximize the dissymmetry factor at a given angle (see Fig. \ref{fig:4} for $\beta=3^\circ$) are similar to the ones that maximize the enantio-sensitive response upon spatial averaging.

\end{document}